\documentclass{article}
\usepackage{arxiv,graphicx,amsmath,float,amssymb}
\usepackage[version=3]{mhchem}
\usepackage{mathtools}
\usepackage{physics}
\usepackage[dvipsnames]{xcolor}
\usepackage{dcolumn}
\usepackage[utf8]{inputenc}
\usepackage[T1]{fontenc}
\usepackage{etoolbox}
\usepackage{arxiv}
\usepackage{hyperref}       
\usepackage{doi}

\title{Fisher Information for Smart Sampling in Time-Domain Spectroscopy}

\author{ \href{https://orcid.org/0000-0003-0893-5743}{\includegraphics[scale=0.06]{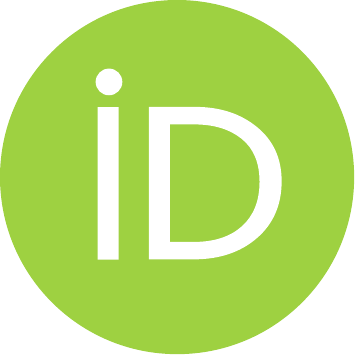}\hspace{1mm}Luca Bolzonello} \\
	ICFO - Institut de Ciencies Fotoniques\\
	The Barcelona Institute of Science and Technology\\
	Castelldefels, Barcelona, 08860 Spain \\
	\texttt{luca.bolzonello@icfo.eu} \\
	\And
 \href{https://orcid.org/0000-0003-4630-1776}{\includegraphics[scale=0.06]{orcid.pdf}\hspace{1mm} Niek F. van Hulst} \\
	ICFO - Institut de Ciencies Fotoniques\\
	The Barcelona Institute of Science and Technology\\
	Castelldefels, Barcelona, 08860 Spain \\
	\texttt{Niek.vanHulst@icfo.eu} \\
 \And
	\href{https://orcid.org/0000-0002-2156-6973}{\includegraphics[scale=0.06]{orcid.pdf}\hspace{1mm}Andreas Jakobsson} \\
	Centre for Mathematical Sciences \\
	Lund University\\
	Lund SE-221 00, Sweden \\
	\texttt{andreas.jakobsson@matstat.lu.se} \\
}

\renewcommand{\headeright}{}
\renewcommand{\undertitle}{}
\renewcommand{\shorttitle}{Fisher Information in Time-Domain Spectroscopy}

\hypersetup{
pdftitle={Fisher Information in Time-Domain spectroscopy},
pdfsubject={Fisher Information in Time-Domain spectroscopy},
pdfauthor={Luca Bolzonello, Niek F. van Hulst, Andreas Jakobsson},
pdfkeywords={Optimal sampling, Time-domain spectroscopy, 2DES, Fisher information, compressed sensing},
}

\begin{document}
\maketitle

\begin{abstract}
Time-domain spectroscopy encompasses a wide range of techniques, such as FTIR, pump-probe, FT-Raman, and 2DES. These methods enable various applications, such as molecule characterization, excited state dynamics studies, or spectral classification. Typically, these techniques rarely use sampling schemes that exploit the {\em prior} knowledge scientists typically have before the actual experiment. Indeed, not all sampling coordinates carry the same amount of information, and a careful selection of the sampling points may notably affect the resulting performance.
In this work, we rationalize, with examples, the various advantages of using an optimal sampling scheme tailored to the specific experimental characteristics and/or expected results. We show that using a sampling scheme optimizing the Fisher information minimizes the variance of the desired parameters. This can greatly improve, for example, spectral classifications and  multidimensional spectroscopy. We demonstrate how smart sampling may reduce the acquisition time of an experiment by one to two orders of magnitude, while still providing a similar level of information.

\end{abstract}

\section{Introduction}

Optical spectroscopy in the time domain and optimal design of experiments are two topics that are rarely connected.
The former is a field that includes all forms of measurements collected in the time coordinate, where time can be considered either the real time of the experiments or the delay between pulses or interferometric branches. It includes both the Fourier-transform (FT-) spectroscopies, such as Fourier-transform Infrared spectroscopy (FTIR)\cite{Berthomieu2009FourierSpectroscopy,Fernandez2018CompressedNano-imaging}, Fourier-transform Raman spectroscopy (FT-Raman)\cite{Hirschfeld1986FT-RamanJustification,Naumann2001FT-INFRAREDRESEARCH},  and the time-resolved spectroscopies such as fluorescence lifetime\cite{Millar1996Time-resolvedSpectroscopy,Suhling2005Time-resolvedMicroscopy,Biskup2014FluorescenceImaging}, pump-probe\cite{Whittock2022ASunscreens,Fushitani2008ApplicationsSpectroscopy}, two dimensional vibrational\cite{Fayer2013VibrationalSpectroscopy,Khalil2003CoherentSolution} or electronic\cite{Gelzinis2019Two-dimensionalNon-specialists,Collini20212DApplications} spectroscopies (2DIR, 2DES). All these techniques find applications, for example, in material characterization, molecular classification, and the study of excited state dynamics.
On the other hand, optimal design of experiments is an area of statistics that aims to optimize the sampling scheme of an experiment, maximizing the information for the estimation of chosen parameters\cite{Atkinson2007OptimumSas,Shin2016NonadaptiveRegression,Goos2011OptimalBooks}. 
In a spectroscopic measurement, these parameters take the form of the frequencies, amplitudes, and widths of relevant spectral peaks.

Why is it then important to connect these fields? As the goal of spectroscopy is to accurately measure relevant parameters, the optimal design of the experiment offers a way to find the sampling coordinates carrying most of the information, avoiding the acquisition of noisy data points with low information content. This approach can vastly reduces the acquisition time with all the related advantages. For example, it allows the saving of photons in experiments with a low photon budget, avoiding sample degradation, and reducing duration of time-consuming experiments. 

How can one then optimize a sampling scheme in the time domain? If one has no idea at all about what to look for and what to expect as a signal, then the prior information is uniformly distributed in the time-space and any collected point can be expected to carry the same amount of information as any other. In this case, a spectroscopist has to iteratively try different time ranges and steps to find where the features of the signal are to be found. However, this is rarely the case in spectroscopy, and most of the time a spectroscopist has a substantial level of prior knowledge about the experiment.

In this work, we rationalize how to exploit this foreknowledge and, based on that, how to optimize the sampling scheme to get most of the information with minimum number of acquisitions.
Spectroscopists generally do some instinctive (automatic) sampling optimization, such as choosing the time step and the time range in oscillating and/or decaying signals. The time step has to comply with the Nyquist boundaries of the bandwidth one is looking for, while the time range determines the frequency resolution in the oscillating signal; there is little reason to exceed the time where the signal has decayed far below the noise level.
These trivial rules are taken for granted in spectroscopy while they are all based on the fact that one knows something in advance, for example, the expected lifetime or bandwidth. 

In oscillating signals, uniformly distributed sampling in the time domain is convenient for the application of fast Fourier transform algorithms (FFT), as usually the desired observables (peak height, width, and so on) are defined in the frequency domain. However, the information concerning these observables is generally not uniformly distributed in the time domain. Consequently, an optimal sampling is likely not uniform\cite{Palmer2014PerformanceNMR,Schuyler2011Knowledge-basedNMR}. A simple example can be represented by a free induction decay (FID) or, more generally, a decaying oscillating signal, where most of the information will reside at the beginning of the measurement, where the signal-to-noise ratio (SNR) is notably higher. Thus, a sampling scheme that is denser at earlier times and sparser at longer times will collect more information than a uniform scheme for the same number of acquisitions. This is the idea behind non-uniform gap sampling, such as deterministic gap\cite{Worley2015DeterministicSampling} or Poisson gap sampling\cite{G.Hyberts2010Poisson-GapData}, already shown to be very efficient in characterizing FID in NMR spectroscopy\cite{Kasprzak2021ClusteredSampling}, where the time steps are increasing along the time axis and, at the same time, following a Poisson distribution.
This leads to the need for non-uniform spectral density estimation algorithms, to reconstruct the signal. This can include non-uniform discrete Fourier transform\cite{Bagchi1999TheProcessing,Duijndam1999NonuniformTransform} but also compressed sensing (CS) methods\cite{Donoho2006CompressedSensing}, that can make use basis changes\cite{Roeding2017OptimizingSpectroscopy}, least squares (LS) estimations with or without Lasso or Ridge regularizations\cite{Andries2013SparsePerspective}, sparse exponential mode analysis\cite{Wang2020CompressedSpectra, Carlstrom2019RapidMethod,Sward2018DesigningData, Sward2016HighSignals}, and more\cite{Potts2016EfficientFFT,Elsener2019SparseMeasurements}. 
The method we propose in the following sections, does not aim to reconstruct the signal, as in CS, but more to retrieve very specific information, based on a certain level of foreknowledge. Through some typical case studies, we demonstrate how to translate the {\em a priori} knowledge into a model function, or a set of functions, that is used to retrieve the optimal sampling, maximizing the accuracy of the desired parameters.

\section{Information, noise, variances and observations in time domain}\label{Info}

Before progressing further in our discussion, it is essential to clarify a fundamental concept: the quantity of information is inversely related to the degree of uncertainty, represented by the noise power, $\sigma^2$. This relationship underscores the principle that measurements are conducted to diminish the uncertainty associated with a parameter, \( \theta \), that we aim to estimate from observations, \( o \).
Each measurement obtained in an experimental or observational context is inherently affected by noise or uncertainty. This noise introduces variability into our observations, which, in turn, influences the precision with which parameters can be estimated. When observations are direct measurements of the parameter of interest, the uncertainty in the measurement directly translates to uncertainty in the parameter estimate $\sigma_\theta^2 \equiv \sigma_o^2$, such as is the case when estimating the intensity of light by detecting the intensity of light itself. However, for parameters derived from complex models, such as decay constants in time-resolved experiments, the connection between the variance of the parameter estimates (\(\sigma_\theta^2\)) and the variances of the observations (\(\sigma_o^2(t_n)\)) at different times \(t_n\) becomes more complicated.

To analytically relate the observed data point variance to the variance of the estimated parameters, we utilize statistical tools that quantify how much information each observation contributes towards the quality of the  parameter estimates. This is achieved using Fisher information matrix\cite{Ly2017AInformation} (FIM), \( \mathcal{I}\), defined as
\begin{align}
\mathcal{I} = - E\left[ \begin{array}{ccc}
\frac{\partial^2 \log L}{\partial \theta_i^2} & \cdots & \frac{\partial^2 \log L}{\partial \theta_i \partial \theta_j} \\
\vdots & \ddots & \vdots \\
\frac{\partial^2 \log L}{\partial \theta_j \partial \theta_i} & \cdots & \frac{\partial^2 \log L}{\partial \theta_j^2}
\end{array} \right] \label{eq:hess}
\end{align}
where $E$ denotes the statistical expectation and \( L\) the likelihood function of the parameters given the observation. This likelihood function is defined via the noise distribution, as \( L\) is by definition the probability of measuring $o$, assuming the parameters \( \theta \).

\subsection{Noise in spectroscopy}
In spectroscopy, where signals are often collected through light detectors, an intrinsic form of noise encountered is shot noise, which is inherently Poissonian due to the discrete nature of photon detection\cite{Lachs1968EffectsDetection}. 
However, the total noise of an experiment does not arise solely from shot noise. It typically includes contributions from various sources, including instrumental and thermal noise, which is prominent in electronic systems. Each type of noise has its characteristic distribution and origin. According to the Central Limit Theorem, when multiple independent noise sources contribute to the overall noise, their combined effect tends toward a normal (Gaussian) distribution\cite{Hoeffding1994AsymptoticNormality}. This theorem applies regardless of the individual distributions of the noise sources, provided there are enough contributing sources or noise events.
Thus, in practical spectroscopy, the aggregate noise can reasonably be modeled as Gaussian. For this reason, we will here consider the noise to be well modelled as being normal distributed, defining the likelihood function as

\begin{equation}
    L(\theta; o) = \prod_n \frac{1}{\sqrt{2\pi \sigma_o^2(t_n)}} \exp\left(-\frac{(o(t_n) - f(t_n, \theta))^2}{2\sigma_o^2(t_n)}\right)
\end{equation}

where $f(t_n, \theta)$ is the model function.

In our simulations, we assume a constant Gaussian noise level over time. However, it is possible to set arbitrarily the noise level $\sigma_o^2(t_n$) for every $t_n$. There are, indeed, at least two cases where noise exhibits particular time or function dependencies, that should be mentioned.

The first situation is the case of pure shot noise that has the typical Poisson distribution. If the number of photons detected exceeds about 20-30 per observation, the Poisson distribution can still be well-approximated by a Gaussian distribution, with the variance equal to the mean. It would be then sufficient for a correct analysis to set the $\sigma_o^2(t_n)$ equal to the function evaluation $f(t_n)$ for every $ t_n$.
The second condition is the case of phase noise, which is linked to uncertainties in the time coordinates \( t_n \). The influence of phase noise \( \sigma_{t_n}^2 \) on the observation variance \( \sigma_o^2(t_n) \) can be determined through the error propagation
\begin{equation}
\sigma_o^2(t_n) \approx (f'(t_n))^2 \sigma_{t_n}^2
\end{equation}
where $f'(t_n)$ is the first derivative of the model function. 
The larger the derivative at a particular \( t_n \), the larger the measurement error induced by any displacement in \( t_n \). If the phase noise is non-negligible with respect to the other noise sources, it should be incorporated, giving a model with a variable variance along the time domain.

\subsection{Example: Simple exponential signal}

As an illustrative example, consider a signal of interest formed by a single exponential decay model, such that
\begin{equation}
 f(t) = a e^{-\gamma t }
 \label{eq:exp}
\end{equation}
where $a$ and $\gamma$ denote the amplitude and decay rate of the exponential, respectively.
In order to estimate the unknown parameters, \(a\) and \(\gamma\), one generally measures the signal decay at various time points, \(t_n\). Each measurement yields an observation \(o(t_n)\) affected by noise,
\begin{equation}
o(t_n) = a e^{-\gamma t_n} + \epsilon_n
\end{equation}
where the additive noise \(\epsilon_n\) is assumed to be well modelled as being normally distributed noise with zero mean and variance \(\sigma_o^2(t_n)\).
The task at hand in this example is thus to estimate the parameters \(\gamma\) and \(a\), which may be done using methods such as LS or maximum likelihood. However, before doing any measurement, one may already estimate the uncertainties \(\sigma_\gamma^2\) and \(\sigma_a^2\) just knowing \(\sigma_o^2(t_n)\) and the \(t_n\) vector. The minimal achievable variance of these estimates can be determined using the Cramér-Rao Lower Bound (CRLB), which is formed as the inverse of the FIM. 

Forming the logarithm of the likelihood function, yields  
a sum of contributions from each sampling point $n$, i.e.,
\begin{align}
\log L(a, \gamma; o) &= - \sum_n \frac{(o(t_n) - a e^{-\gamma t_n})^2}{2\sigma_o^2(t_n)}  \nonumber \\
& \quad\quad\quad -\sum_n \frac{1}{2} \log(2\pi \sigma_o^2(t_n))
\label{logL}
\end{align} 
Curiously, maximizing this log-likelihood corresponds to minimizing the sum of squared residuals for \(a\) and \(\gamma\), as the only element of the log-likelihood varying with these parameters is $\sum_n (o(t_n) - a e^{-\gamma t_n})^2$, showing the equivalence between the least square and maximum likelihood estimates in the case of white Gaussian noise.
The FIM for the parameters \( a \)  and \( \gamma \) in this model may be calculated by combining \eqref{eq:hess} and \eqref{logL}, such that
\begin{align}
\mathcal{I} &=-E
\begin{bmatrix}
\frac{\partial^2 \log L}{\partial a^2} & \frac{\partial^2 \log L}{\partial a \partial \gamma} \\
\frac{\partial^2 \log L}{\partial \gamma \partial a} & \frac{\partial^2 \log L}{\partial \gamma^2}
\end{bmatrix}
\nonumber \\
&=\begin{bmatrix}
\sum_n \frac{e^{-2\gamma t_n}}{\sigma_o^2(t_n)} & \sum_n \frac{a t_n e^{-2\gamma t_n}}{\sigma_o^2(t_n)} \nonumber \\
\sum_n\frac{a t_n e^{-2\gamma t_n}}{\sigma_o^2(t_n)} & \sum_n \frac{a^2 t_n^2 e^{-2\gamma t_n}}{\sigma_o^2(t_n)}
\end{bmatrix}\\
&=\sum_n\frac{1}{\sigma_o^2(t_n)}\begin{bmatrix}  e^{-2\gamma t_n} &  a t_n e^{-2\gamma t_n} \\ a t_n e^{-2\gamma t_n} &  a^2 t_n^2 e^{-2\gamma t_n} \end{bmatrix}_n
\label{FIM}
\end{align}
The CRLB, which is formed as the inverse of $\mathcal{I}$, determines the minimal variance of any unbiased estimator, i.e., the variance of the sought parameters will be, at least, as large as the inverse of the FIM. Thus, the covariance matrix of the estimated parameters, \( \operatorname{Cov}(\hat{a}, \hat{\gamma}) \), is bounded by
\begin{equation}
\operatorname{Cov}(\hat{a}, \hat{\gamma}) = 
 \begin{bmatrix} \sigma_{\hat{a}}^2 & \sigma_{\hat{a}\hat{\gamma}} \\ \sigma_{\hat{\gamma}\hat{a}} & \sigma_{\hat{\gamma}}^2 \end{bmatrix} \geq \mathcal{I}^{-1}
\end{equation}
where $A \ge B$ indicates that the matrix $A-B$ is positive semidefinite, with \( \sigma_{\hat{a}}^2 \) and \( \sigma_{\hat{\gamma}}^2 \) denoting the variances of the estimators \( \hat{a} \) and \( \hat{\gamma} \), respectively, and \( \sigma_{\hat{a}\hat{\gamma}} = \sigma_{\hat{\gamma}\hat{a}} \) their cross-covariances.
Notably, if the noise power is constant along the time coordinate, such that $\sigma_o^2(t_n)=\sigma_o^2$, one may simplify the FIM, extracting $\sigma_o^2$ from the matrix elements, such that 
\begin{equation}
   \begin{bmatrix} \sigma_{\hat{a}}^2 & \sigma_{\hat{a}\hat{\gamma}} \\ \sigma_{\hat{\gamma}\hat{a}} & \sigma_{\hat{\gamma}}^2 \end{bmatrix} \geq
    \sigma_o^2 \left[\sum_n \begin{bmatrix}  e^{-2\gamma t_n} &  a t_n e^{-2\gamma t_n} \\ a t_n e^{-2\gamma t_n} &  a^2 t_n^2 e^{-2\gamma t_n} \end{bmatrix}_n\right]^{-1}
\end{equation}
This allows for a direct relation between the variances of parameter estimates,  $\sigma_a^2$ and $\sigma_\gamma^2$, and the noise variance  in the measurement $\sigma_o^2$ and the sampling points $t_n$. 

\begin{figure}[h]
\centering
\includegraphics[width=0.45\textwidth]{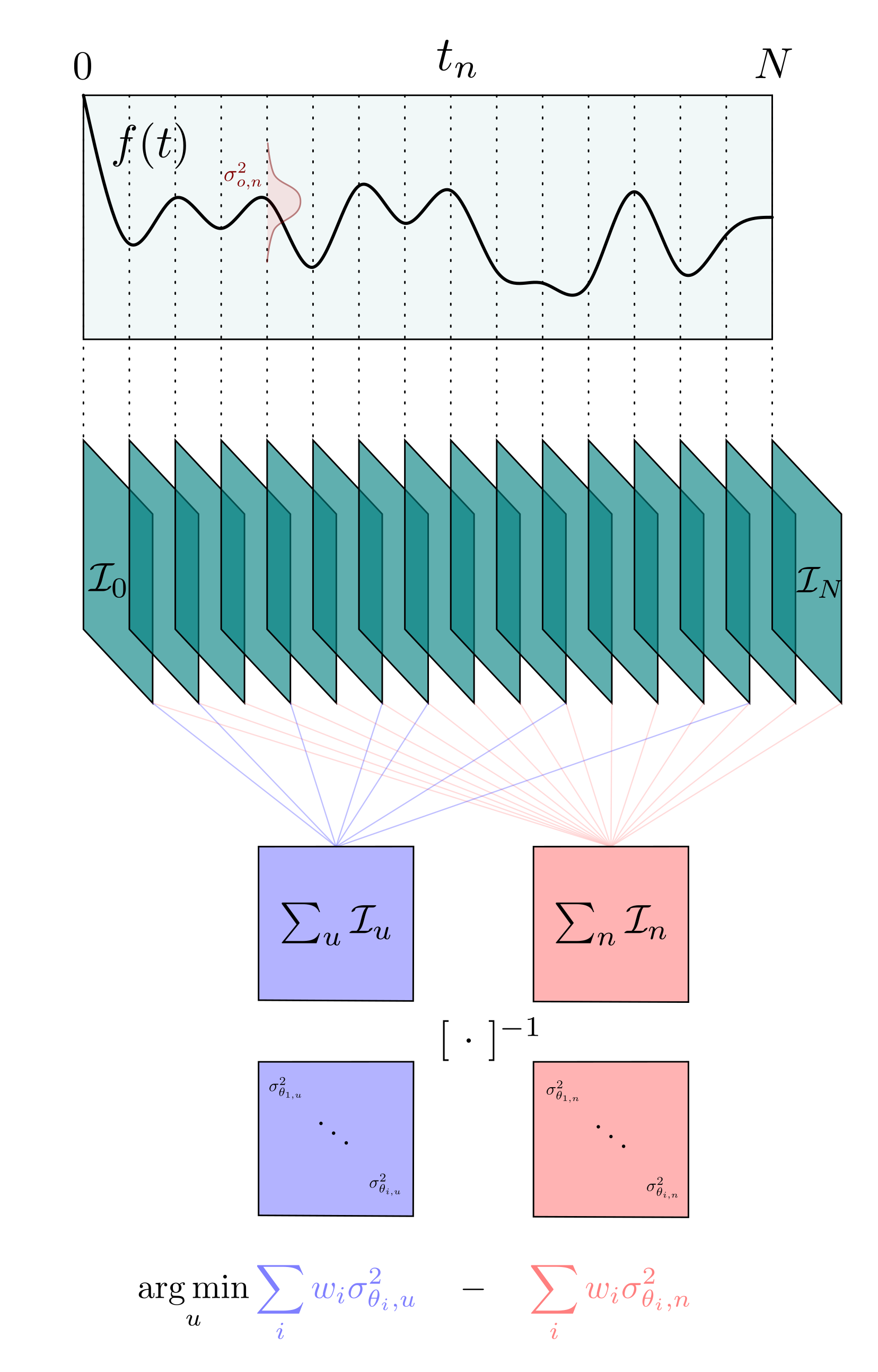}
\caption{An illustration of the introduced Fisher method. An assumed signal is discretized using a uniform grid respecting the Nyquist conditions. For every sampling point $n$,  $\mathcal{I}$ of the parameters of the function $f(t)$ is calculated. As the Fisher information is additive, the sum of all matrices can be combined to a \(\sum_n\mathcal{I}_n\). The inverse of total $\mathcal{I}$ contains on the diagonal the variances of all the parameters, which correspond to the lower bounds for their estimations.  The same procedure can be applied on a subset of sampling points $u$, in this way obtaining a lower bound on the variances of the parameters that could be obtained by acquiring only data on the reduced subset. The variances of the total ($\sum_n$) and selected ($\sum_u$) sampling schemes can be compared to check the goodness of the selected subset.  The minimization of the weighted variances is used to find the best subset $u$.}
\label{fig:fish}
\end{figure}

This relationship indicates that the minimal variances of the parameter estimates are inversely proportional to the accumulated Fisher information, adjusted for the noise level, which quantifies how each observation contributes to our knowledge about each parameter. This method not only enhances our understanding of how to form an appropriate experimental design, but also informs  the minimum variances obtainable with such a measurement, before any measurements has been taken.
Figure~\ref{fig:fish} illustrates the overall procedure.

\subsection{Optimization of sampling for Variances Minimization}

\subsubsection{Minimization Parameter}
Now that the relationship between the sampling points \( n \) and the covariance of the estimators \( \sigma_{\hat{\theta}}^2 \) has been established, our objective shifts to optimizing a sampling scheme aimed at achieving minimum variance for the desired parameters.

There are several criteria to consider when optimizing sampling using the FIM, each suited to different experimental needs and objectives. Among them are D-optimality, which maximizes the determinant of \(\mathcal{I}\); C-optimality, which is more tailored to emphasizing specific contrasts between parameters; and A-optimality, which seeks to minimize the trace of the inverse of the FIM, \( \text{tr}(\mathcal{I}(\theta)^{-1}) \). This corresponds to minimizing the average variance across all parameters.
Given the nonlinear nature of our model and the potential for parameters to vary significantly in magnitude, a simple average (as in standard A-optimality) may not suffice. To address this, we can consider a weighted version of A-optimality, where weights are applied to the trace of \(\mathcal{I}^{-1}\) based on the relative importance or scale of the parameters, such that
\begin{equation}
W = \text{diag}(w_i) \cdot \text{tr}\left[\mathcal{I}(\theta)^{-1}\right] = \sum_i w_i \sigma_{\theta_i}^2   
\end{equation}
where \( W \) is the overall cost function to minimize, with \( w_i \) representing weights that could be chosen based on the relative importance of each parameter or based on their expected magnitudes. This ensures that parameters with greater expected variances or more significance impacts on the model's outcome are given appropriate attention in the design.
In summary, the aim to design the experiment in such a way that the weighted trace of \(\mathcal{I}^{-1}\), \( W \), is minimized.

\subsubsection{Sampling Selection method}

As noted, minimizing \(W\) is our primary goal, but it is important to recognize that simply adding sampling points will reduce the minimal \(W\), as more points generally increase the information, thereby reducing the variances. However, our objective is to optimize these points while considering a limited number of samples at our disposal. The method employed involves generating an over-sampled grid of time points, \(t_n\) with \(N\) samples, and then selecting a subset of these, denoted \(u\) aiming to minimize the difference between the variance captured by subset and the full sampling\cite{Sward2018DesigningData}, \(W_u - W_N\). This optimization is a convex function over the selected subsamples and can be effectively minimized using convex minimization algorithms.
\begin{figure}[h]
\centering
\includegraphics[width=0.45\textwidth]{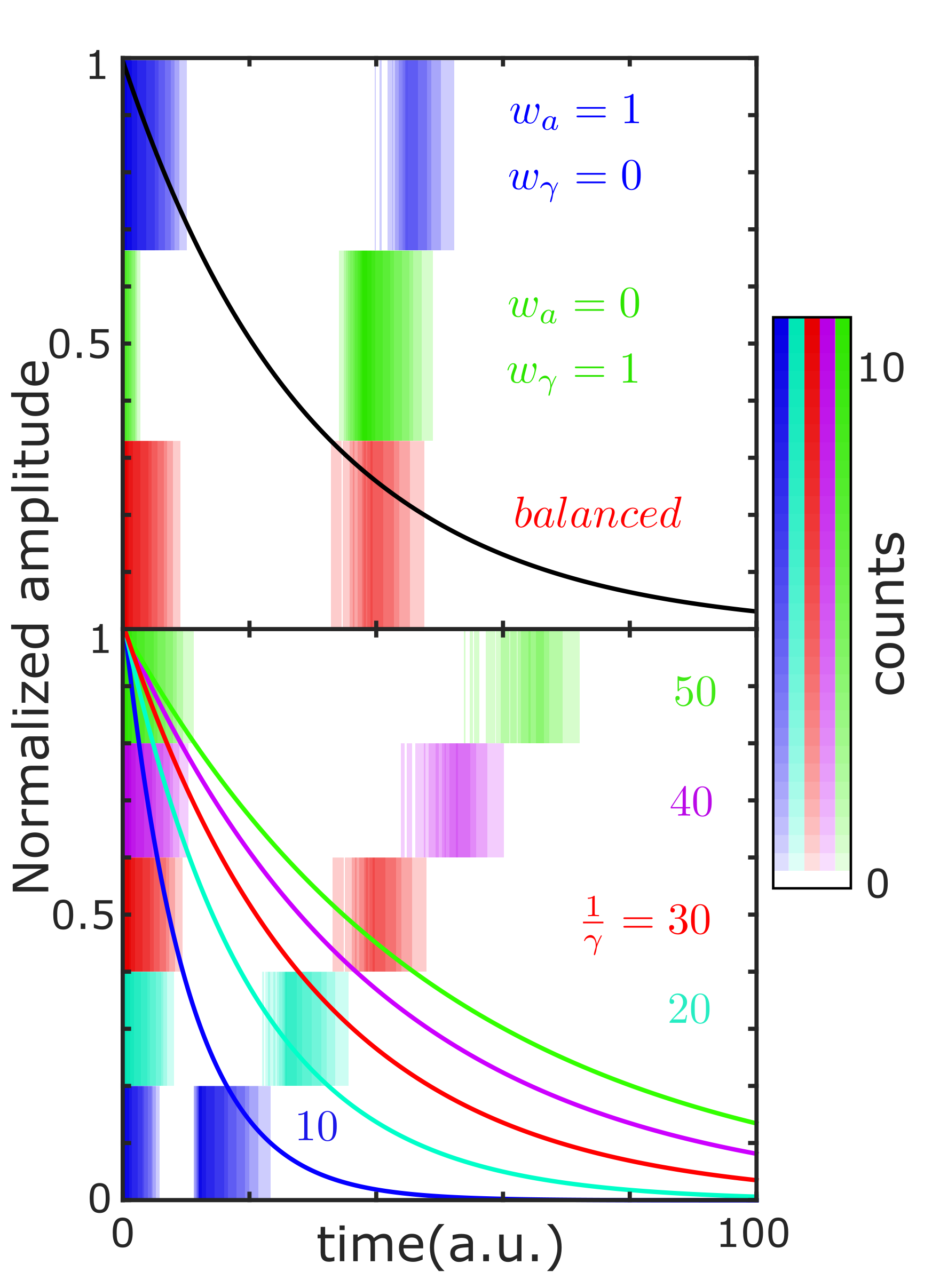}
\caption{Optimal sampling for exponential function, $ae^{-\gamma t}$. Top graph shows the best sampling varying the weights between parameters. The information is displayed as heatmap in the background. The best sampling when the minimization run only on the amplitude parameter is shown in blue, only for the decay rate in green and when they are balanced in red. Bottom graph shows how the balanced information varies as a function  of the decay rate of the model function.}
\label{fig:decays}
\end{figure}
The optimization criterion for the selection of time points of the subset \(u\) as compared to the full set \(N\) may then be expressed as
\begin{align}
  \operatorname*{arg\,min}_u (W_u - W_N) &= \operatorname*{arg\,min}_u \Bigl(\text{diag}(w_i) \cdot \text{tr}[\mathcal{I}_u(\theta)^{-1}] \Bigr. \nonumber \\
  & \quad \quad \quad \Bigl. - \text{diag}(w_i) \cdot \text{tr}[\mathcal{I}_N(\theta)^{-1}] \Bigr) \nonumber \\
  & \hspace{-10mm} = \operatorname*{arg\,min}_u \left(\sum_i w_i (\sigma_{\theta_i,u}^2 - \sigma_{\theta_i,N}^2)\right)
\end{align}
Figure~\ref{fig:decays} shows the information distribution for the sought parameters \(\begin{bmatrix}
    a&\gamma
\end{bmatrix}^T\) of the exponential function \(ae^{-\gamma t} \)for varying weights  \(\begin{bmatrix}
    w_a&w_\gamma
\end{bmatrix}\). This distribution is obtained by evaluating the exponential function on a full grid of 1001 points between 0 and 100, and counting the number of times a specific $t_n$ is selected when varying the number of elements in the subset, i.e., when selecting, for example, 1\%, 2\%, or 3\%, accumulating the number of times a given sample has been selected in each simulation, we obtain the information heatmap shown behind the graphs. As can be seen in the figure, the information in all cases is focused on 2 areas, one close to $t=0$  and the other around $t\approx\frac{1.2}{\gamma}$. When the focus is on minimizing the variance of $a$, the information is more focused in the initial part of the signal, whereas when the focus is on minimizing the variance of $\gamma$, the information is more focused in the latter part of the signal. This makes intuitive sense; the value of $a$ is easiest to determine in the initial part of the signal when this has an high SNR, so the more points that are selected at the beginning of the signal, the more precise the estimate of $a$ will be. However, not all samples should be selected there, as one also needs to determine the decay rate to estimate $a$ properly, so there will be always the need to measure in the second area. Similarly, when instead focusing the attention on minimizing the variance of $\gamma$, it becomes more important to determine the decay rate properly, and one then needs more information later on in the signal. 

\section{Applications}

\subsection{Exploiting the prior knowledge}

Until now, our discussion has focused on minimizing the variance of estimators for predefined parameters within our model, assuming a clear prior knowledge of both the model and its parameters' values. This approach is ideal for controlled scenarios, such as quantifying or classifying different but known molecules present in the same sample.
To illustrate a more practical use of the method, we examine several case studies in spectroscopy. Initially, we consider scenarios involving simple exponential decays, commonly found in fluorescence lifetime techniques. Then, we proceed to explain how to treat oscillating signals, both real- and complex-valued, that can be used in Fourier transformed techniques. We illustrate this case by showing an example of sampling optimization in molecule classification through FT-raman.
Finally, we examine a case in which the prior knowledge is limited, as so often happens in real-world experiments. We illustrate this case with an example of Action-Two-Dimensional Electronic Spectroscopy, examining how one may retrieve an optimal sampling scheme using a Monte Carlo approach to probe a broader range of expected parameters' values.

\subsection{Exponential decaying signals}

\begin{figure}[ht]
\centering
\includegraphics[width=0.45\textwidth]{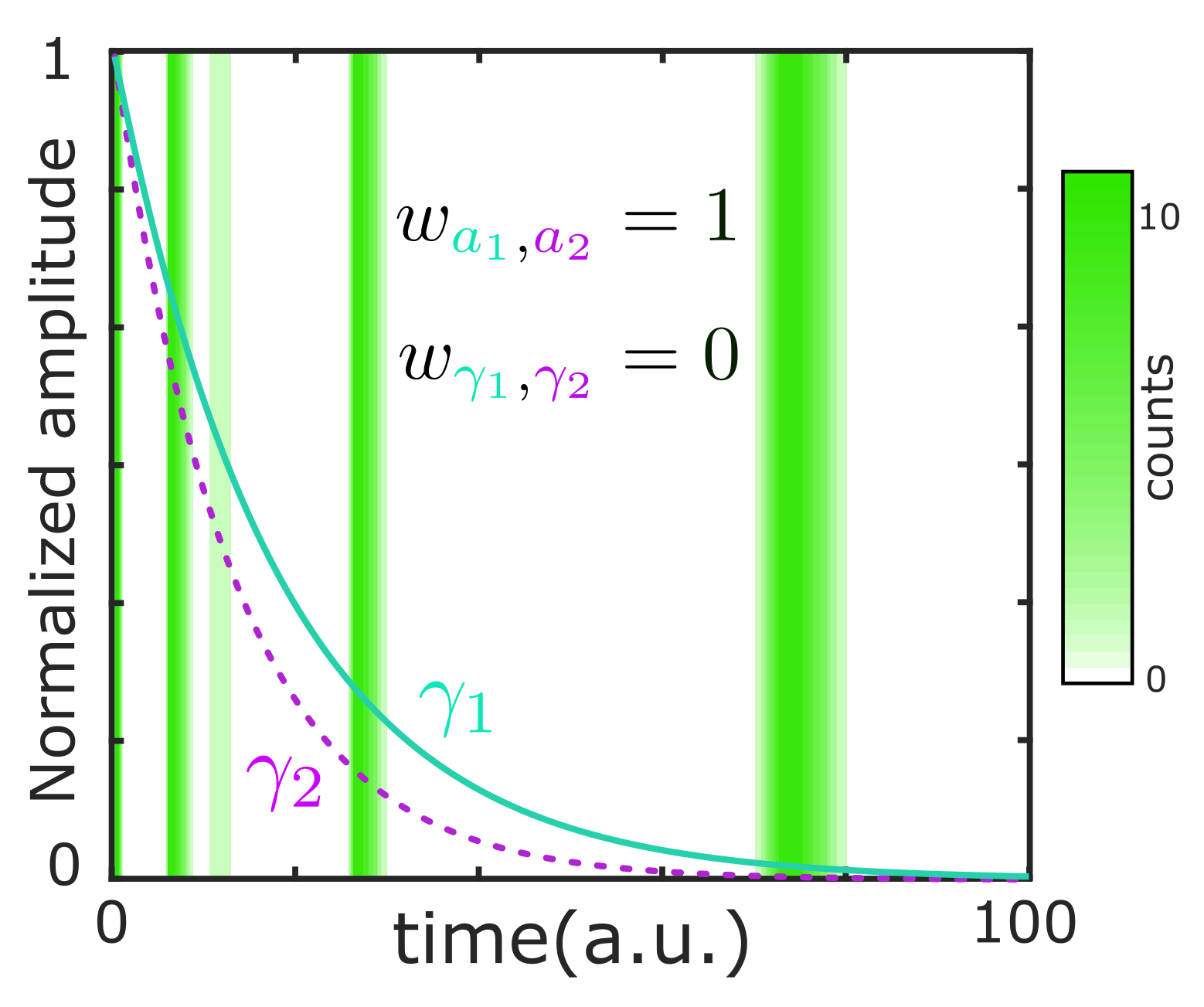}
\caption{Information distribution for estimating the amplitudes of a sum of real exponential functions.}
\label{fig:compdecays}
\end{figure}

One of the most typical time-dependent signals in spectroscopy is the exponential decay. This prevalence is not coincidental; exponential functions are the solutions to differential equations that describe the stochastic decay processes of quantum states.
Here, we focus on exploring scenarios that can be well modelled as a sum of exponentials, as is typical in Fluorescence Lifetime Imaging (FLIM). In FLIM, different decay rates and amplitudes are used to identify the concentrations of specific molecules within a spatial region. While typical FLIM setups, such as those using the Time-Correlated Single Photon Counting (TCSPC) methods, may not be compatible with sampling reduction, other scenarios might allow for specific gating of the signal to optimize data acquisition.
As a general and illustrative example, consider a signal composed of two exponentials, such that
\begin{equation}
f(t) = \sum_{k=1}^2 a_k e^{-\gamma_k t}    
\end{equation}
which is detailed by the unknown parameters
\begin{equation}
\theta = \left[\begin{array}{cccc} a_1 & \gamma_1 & a_2 & \gamma_2 \end{array} \right]^T
\end{equation}
Considering a case when one seek to estimate the signal amplitudes as accuratly as possible, which would be crucial, for example, for quantifying molecular concentrations in FLIM, we select the weighting vector to focus only on these two parameters, i.e.,
\begin{equation}
w_\theta = \left[\begin{array}{cccc} 1 & 0 & 1 & 0 \end{array} \right]
\end{equation}
As shown in Figure~\ref{fig:compdecays}, the information about the amplitudes is not mainly centered close to \(t=0\), as it was for the single exponential. The reason for this is that the optimization algorithm seeks the best sampling strategy to maximize information on both amplitudes simultaneously, thus aiming to sample at time coordinates where the two exponentials exhibit the greatest differences.

\subsection{Oscillating time-domain signals}

\begin{figure}[ht]
\centering
\includegraphics[width=0.45\textwidth]{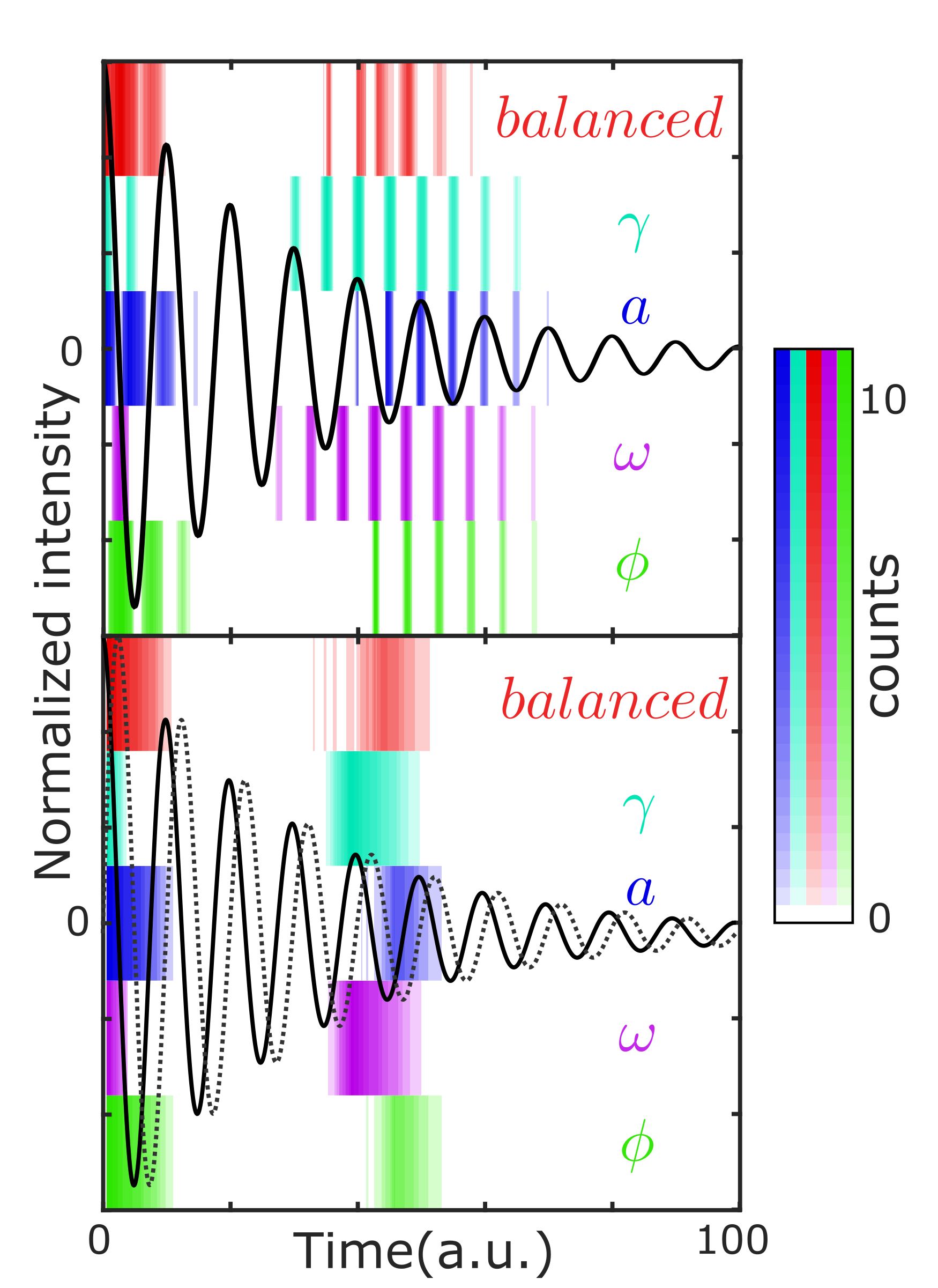}
\caption{Comparison between the information distributions between real-valued acquisition (top graph) and complex-valued acquisition (bottom graph), as a function of different weights.}
\label{fig:realcomplex}
\end{figure}

Another typical spectroscopic signal is characterized by oscillating signals, typically originating from coherences between quantum states. Oscillations measured in the time domain are often analyzed using Fourier Transform (FT) techniques, as the parameters of interest, such as the amplitude and frequency of a peak, or its shape, are typically expressed in the frequency domain. 
This form of signals can be expressed in the time domain using three components: the amplitude \(a\), the oscillating part \(s(t)\), and the lineshape \(L(t)\), such that
\begin{equation}
f(t) = a \cdot s(t) \cdot L(t) 
\end{equation}
The relationship between the time signal and its frequency counterpart is established through the Fourier Transform:
\begin{equation}
    f(t) \xleftrightarrow{\text{FT}} f(\omega)
\end{equation}
Before entering into specifics, it is crucial to highlight a significant distinction in the techniques used for acquiring oscillating signals, particularly their phase sensitivity. In spectroscopic techniques like FTIR, for instance, the detected signals are photons, meaning that the measured signal is real-valued. This detection typically involves measuring the interferometric autocorrelation of the electric field, where what is observed at the detector is the square of the light electric field, i.e., an integer number of photons. In contrast, techniques that involve phase cycling, holography, and other methods capture both the phase and the intensity of the signal at every observation $o$.
In the first case, \(s(t)\) typically takes the form \( \cos(\omega t + \phi) \), while for phase-sensitive acquisition, \(s(t)\rightarrow e^{i(\omega t + \phi)} \). This distinction leads to different approaches in the sampling scheme. Figure~\ref{fig:realcomplex} illustrates these differences, using a simple exponential decay as the lineshape \( L(t) = e^{-\gamma t} \). In the real-valued case, the information is sensitive to the different phases within the oscillation period, especially at the oscillation's maxima, where amplitude and decay rate are most discernible due to better SNR at these points, which define the envelope. In contrast, the information about the frequency and phase is concentrated at the nodes.

This disparity does not occur with complex-valued data, where the maxima and nodes in the real part find their counterparts in the imaginary part. Although sampling optimization is applicable in both scenarios, acquiring complex-valued data is preferable as the sampling depends only on the envelope, not on the oscillating frequency. If complex-valued data acquisition is not feasible, it is often possible, especially if the subset of data is sufficiently large, to apply a non-uniform Hilbert transform to the collected data. If this is the case, sampling optimization can be effectively performed assuming complex-valued functions.

\begin{figure}[h]
\centering
\includegraphics[width=0.45\textwidth]{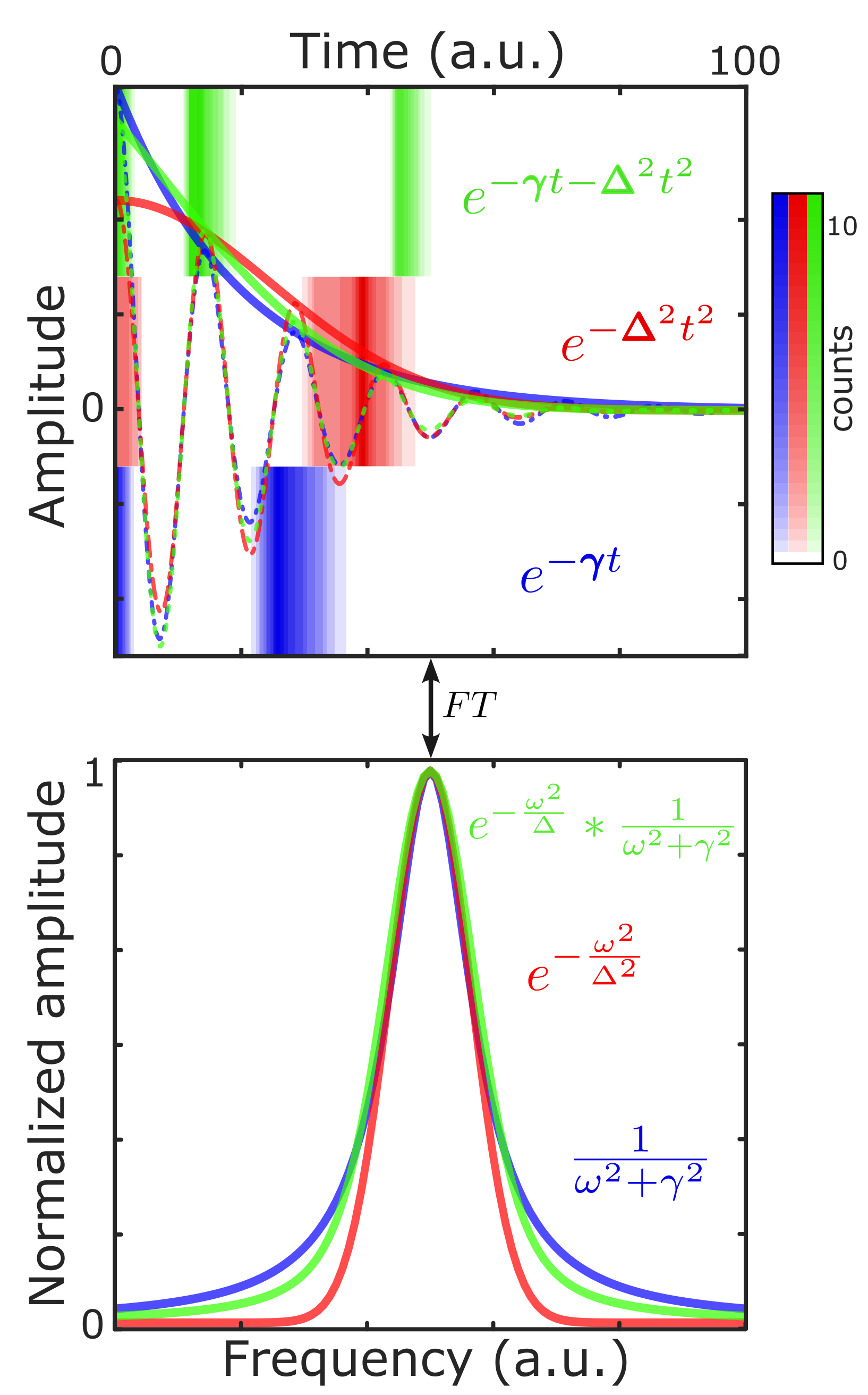}
\caption{The information distribution in the time-domain for the parameters $\gamma$ and $\Delta$, for the three different lineshapes.}
\label{fig:Lineshapes}
\end{figure}

It is worth noting that, for complex-valued oscillating signals, the frequency and phase do not influence the information of the model function, implying what becomes important to target is instead the lineshape, $L(t)$.
The lineshapes of oscillating signal are function that are related to the correlation between the fluctuations of the energies of the quantum states involved in the coherence.
It is common to distinguish in 2 limit cases, when the fluctuations are much faster or much slower than the coherence. The former is called homogeneus broadening and leads to a exponential decay of the oscillation the second inhomogeneous and leads to an Gaussian decay of the oscillation decay.
Their frequency counterparts are a Lorentzian and a Gaussian function, respectively, formed as
\begin{align}
L_h(t) = e^{-\gamma t} &\xleftrightarrow{FT} L_h(\omega)\propto \frac{1}{{\gamma^2}+\omega^2}\\
L_i(t) = e^{-\Delta^2 t^2} &\xleftrightarrow{FT} L_i(\omega)\propto  e^{-\frac{\omega^2}{\Delta^2 }}
\end{align}
As an alternative, a Voigt profile is often used to account when both effect are present simultaneously, using
\begin{equation}
    L_v(t) = e^{-\gamma t}\cdot e^{-\Delta^2 t^2} \xleftrightarrow{FT} L_v(\omega)\propto e^{-\frac{\omega^2}{\Delta^2 }}\ast \frac{1}{{\gamma^2}+\omega^2}\\
\end{equation}
Comparing these lineshapes, one can see how the function profile in frequency is related to the time envelope of the decaying oscillating signals. As a result, one may optimize the sampling in time to maximize the information for the shape parameters, $\Delta$ and $\gamma$.
Figure~\ref{fig:Lineshapes} illustrates the information content for these parameters for the considered example. It should be noticed that while for pure homogeneous or inhomogeneous broadening there are two main areas of information, whereas for the Voigt signal, there are three areas which are required to properly estimate both $\Delta$ and $\gamma$. 

\subsection{Lineshapes in Action-2DES}

The potential of optimized sampling techniques in resolving amplitude, frequency, and lineshapes becomes particularly evident when applied to two-dimensional spectroscopies. In these methods, the need for data points scales quadratically with the time range, making optimized sampling even more beneficial.
In 2DES, one aims to capture the ultrafast non-linear response of a sample by interacting with ultrafast laser pulses. Various 2DES configurations and geometries, each with distinct advantages\cite{Fuller2015ExperimentalSpectroscopy}, are currently employed. Notably, the collinear beam geometry is gaining attention for its characteristic action detection, which is detecting a signal that is proportional to the population of the excited state\cite{Bolzonello2023NonlinearDetection}, like fluorescence\cite{Draeger2017Rapid-scanSpectroscopy} and current\cite{Bolzonello2021Photocurrent-DetectedCells,Karki2014CoherentPhotocell}, opening the possibility to study non-transmitting samples or to integrate 2DES with imaging. These action-detection techniques requires a different approach for extracting non-linear signal components, as these involve scanning the third time interval rather than detecting it in the frequency domain through a spectrometer, as it occurs in non-collinear techniques. This can significantly extend the required acquisition time. On top of that, phase cycling is necessary to retrieve the single components of the non-linear signal, necessitating at least 27 times as many acquisitions as in a regular spectroscopic setup\cite{Tan2008TheorySpectroscopy}.
For this reason, many groups have explored various compressed sensing methods to reconstruct signals with fewer acquisitions\cite{Wang2020CompressedSpectra,Spencer2016MappingSensing,Sanders2012CompressedExperiments}.

The primary aim of these methods is signal reconstruction, determining the relevant frequencies and amplitudes, while preserving the lineshape information. \cite{Bhattacharya2017AcceleratingGIRAF,Humston2019OptimizedSpectra}.
A better approach is to think the other way around: optimizing the sampling scheme to have more informative observables to reconstruct the signal. Attempts in this direction has been made from different groups. In one study, the Brixner group uses a projection of an optimized sparse sample along time coordinates into von Neumann basis \cite{Roeding2017OptimizingSpectroscopy}.  Another approach come from the group of Tiwari\cite{Sahu2023High-sensitivityScan}, where they make use of an non-uniform sampling, with exponentially increasing step size. However, in both cases, the optimization of the sampling scheme is made {\em a posteriori}, i.e. as a function of the actual signal acquired. 
Our approach instead aims to retrieve the best sampling pattern as a function of the prior knowledge, that means without knowing with high precision the fully acquired signal.

\begin{figure*}[ht]
\centering
\includegraphics[width=\textwidth]{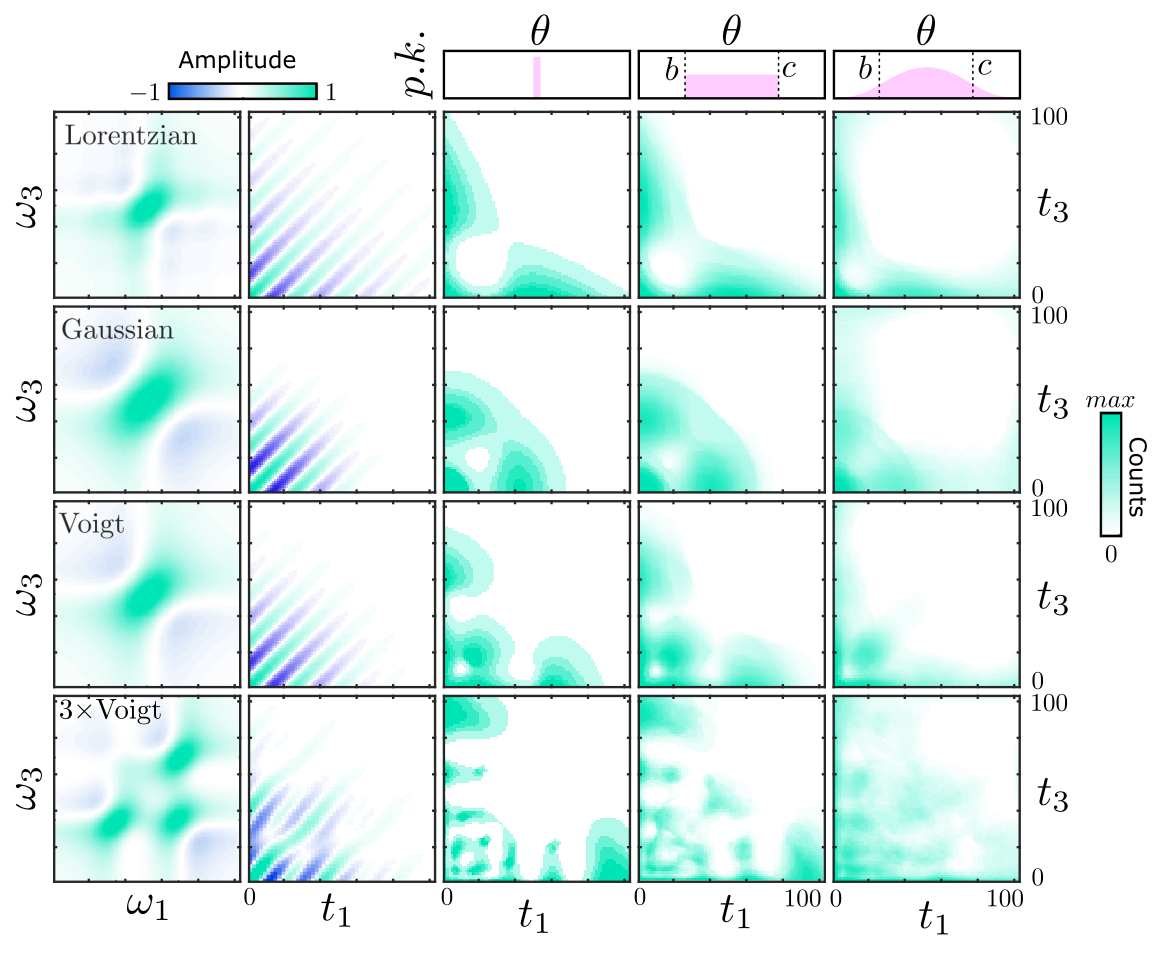}
\caption{The time-domain information distribution for a 2D spectra with varying lineshapes. The first column displays the real parts of the frequency-domain functions with different lineshapes. The second column shows their time-domain counterparts. The third to fifth columns illustrate the lineshape parameters' information distribution, illustrating the case for fixed parameter values, parameters with uniform distribution across a range $[b,c]$, and parameters distributed around an expected value with uncertainty. As can be seen, the information is denser close to the axes for a single peak and more spread across the map with multiple peaks.}
\label{fig4}
\end{figure*}

To select the best acquisition coordinates (\(t_1, t_3\)) that minimize the variance of the parameters for the signal model, one can apply the Fisher information method. In this example, we consider a grid of \(101 \times 101\) points, ensuring the selected parameter values are appropriately captured within this range. Assuming acquisition of complex-valued data, coming from phase cycling, we can model the 2D peaks as 2D Voigt functions, defined as
\[
F(t_1,t_3)= a\cdot e^{-i(\omega_1 t_1 +\omega_3 t_3)}\cdot e^{-\gamma (t_1 + t_3) -\Delta^2( t_1^2 + t_3^2-2t_1 t_3)}
\]
Should one wish to focus on purely Lorentzian lineshapes, set \( \Delta = 0 \); similarly, for a purely Gaussian lineshape set \( \gamma = 0 \).
As observed in the 1D case, when the model function is complex-valued, the optimal sampling points remain unaffected by the value of the frequency parameter. From the third to the fifth column of Figure~\ref{fig4}, this can be seen in the information distribution, shown as a heatmap in the background, where the colormap goes from 0 to the maximum number of coordinate selection. The number of informative regions also varies with the quantity of parameters in the model; more parameters leads to more regions of sampling.
In the third column,  the shown information distribution is for the case when full prior knowledge about the lineshapes parameters, $\gamma$ and $\Delta$, is known. For single-function models, the information is predominantly distributed along the axes. When the model includes multiple peaks, the information disperses more broadly across the map, highlighting that as the complexity of the signal model increases, with more parameters to be estimated, the more the information is spread in the time map.
When the exact values are unknown, a Monte-Carlo simulation has been computed across the considered range of parameters in order to determine a more general optimal sampling pattern. The simulations are computed picking parameters values ($\gamma,\Delta$ or both) using $[b,c]$ range that goes between $\pm 50\%$ of their expected value. In the fourth column the uncertainties is homogeneously distributed along the range while in the fifth one $[b,c]$ act as FWHM of a normal distribution. This approach yields an information distribution that, while somewhat more blurred, retains a degree of localization. This demonstrates that, even without the full knowledge on the final signal, it is possible to reduce the acquisition space by sampling mainly in the most informative areas.

\subsection{Molecules classification in FT-Raman}

\begin{figure}[h]
\centering
\includegraphics[width=0.45\textwidth]{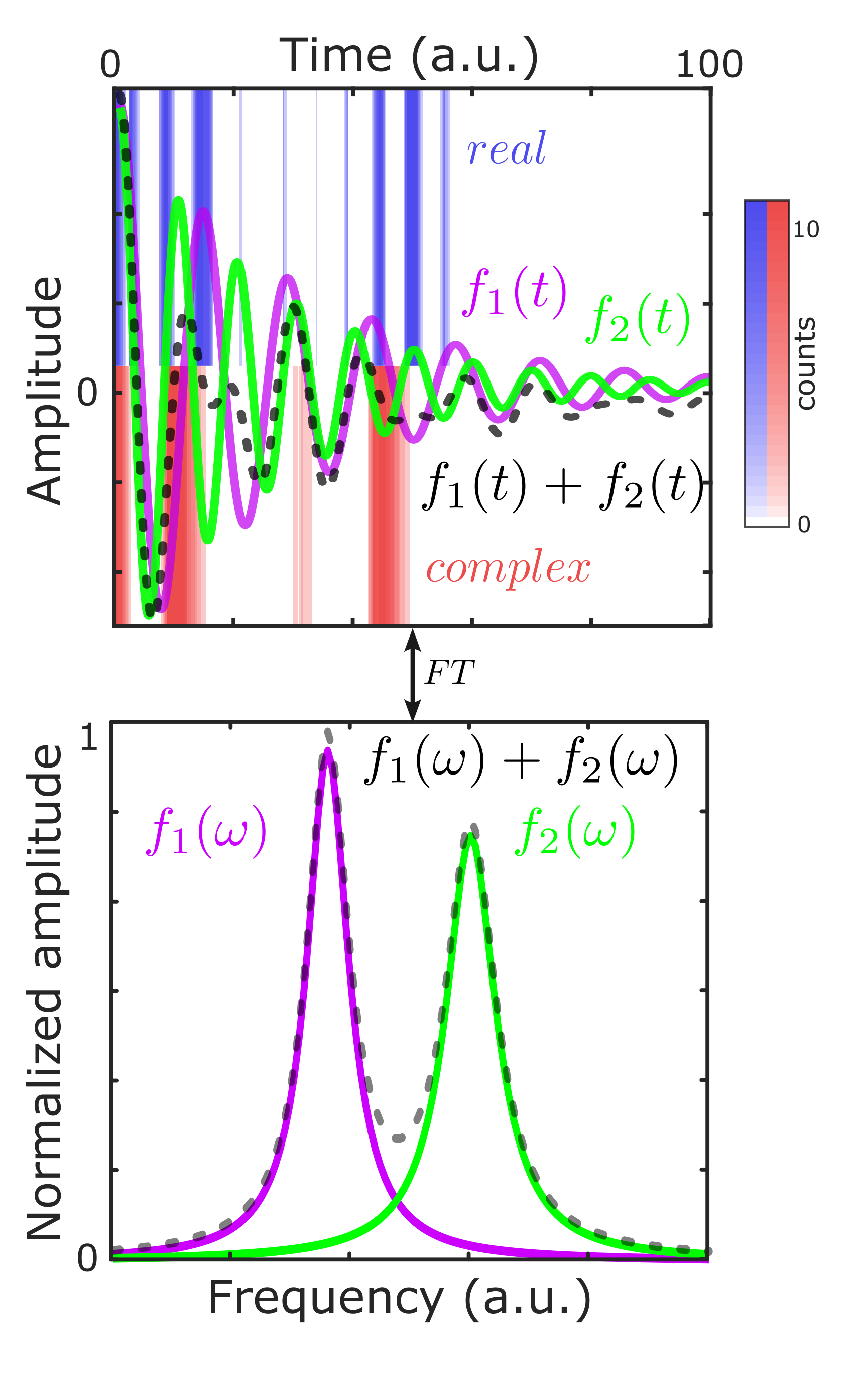}
\caption{Classification through peak amplitudes. Top graph shows optimal sampling pattern for minimizing the variance of amplitude estimates of the sum of two oscillating exponential function, shown for both real- (blue) and complex-valued data (red). Bottom graph, the frequency domain functions and their sum are shown.}
\label{fig:comp exp}
\end{figure}

Minimizing the variance of parameter estimates is crucial for enhancing the classification efficiency of different spectra. Consider the scenario where two peaks in a spectrum correspond to two distinct molecular species. Distinguishing between these species can be effectively achieved by comparing the amplitudes of these peaks. While the linewidth and other spectral features could also serve as discriminators, peak amplitudes are often the most direct and significant indicators.
To optimize the classification for this, we next construct a model function representing the sum of the two peaks. We then determine the optimal sampling that will minimize the variance of each peak's amplitude. This approach improves the precision of amplitude estimation and consequently the overall efficiency of species classification.
It should be noted that, as discussed above, the results may differ between real- and complex-valued data. Figure~\ref{fig:comp exp} illustrates how optimal sampling points are selected to achieve this minimized variance for the amplitudes estimation in both types of data.

We examine the suitable sampling pattern in FT-Raman spectroscopy under realistic conditions. For this reason, we simulate realistic signals, with different numbers of sampling points and noise levels. The simulated frequency domain spectra are generated with five Lorentzian peaks with random amplitudes, frequencies, and linewidths for each molecule. They are then transformed into the time domain, preserving only the real part to emulate a realistic experimental scenario. We chose a full time grid with a step of 0.4~fs from 0 to 1000~fs. This is (with a large margin) within the Nyquist limit for a hypothetical Raman signal generated by a HeNe laser (633~nm / 15790~cm$^{-1}$ / 0.47~fs$^{-1}$). We opted for oversampling the time axis to avoid possible aliasing of stokes and anti-stokes time domain signals that can spread over more than 8000~cm$^{-1}$, and to avoid the possible missing of critical points inside a "bandwidth" period in case of real data acquisition. However, the range of the ratio of the selected points span from 1\% to 12\%, that corresponds to between approximately 5\% and 60\% of the data points that would be necessary according to the Nyquist criterion for the 8000~cm$^{-1}$ bandwidth.

\begin{figure*}[h]
\centering
\includegraphics[width=\textwidth]{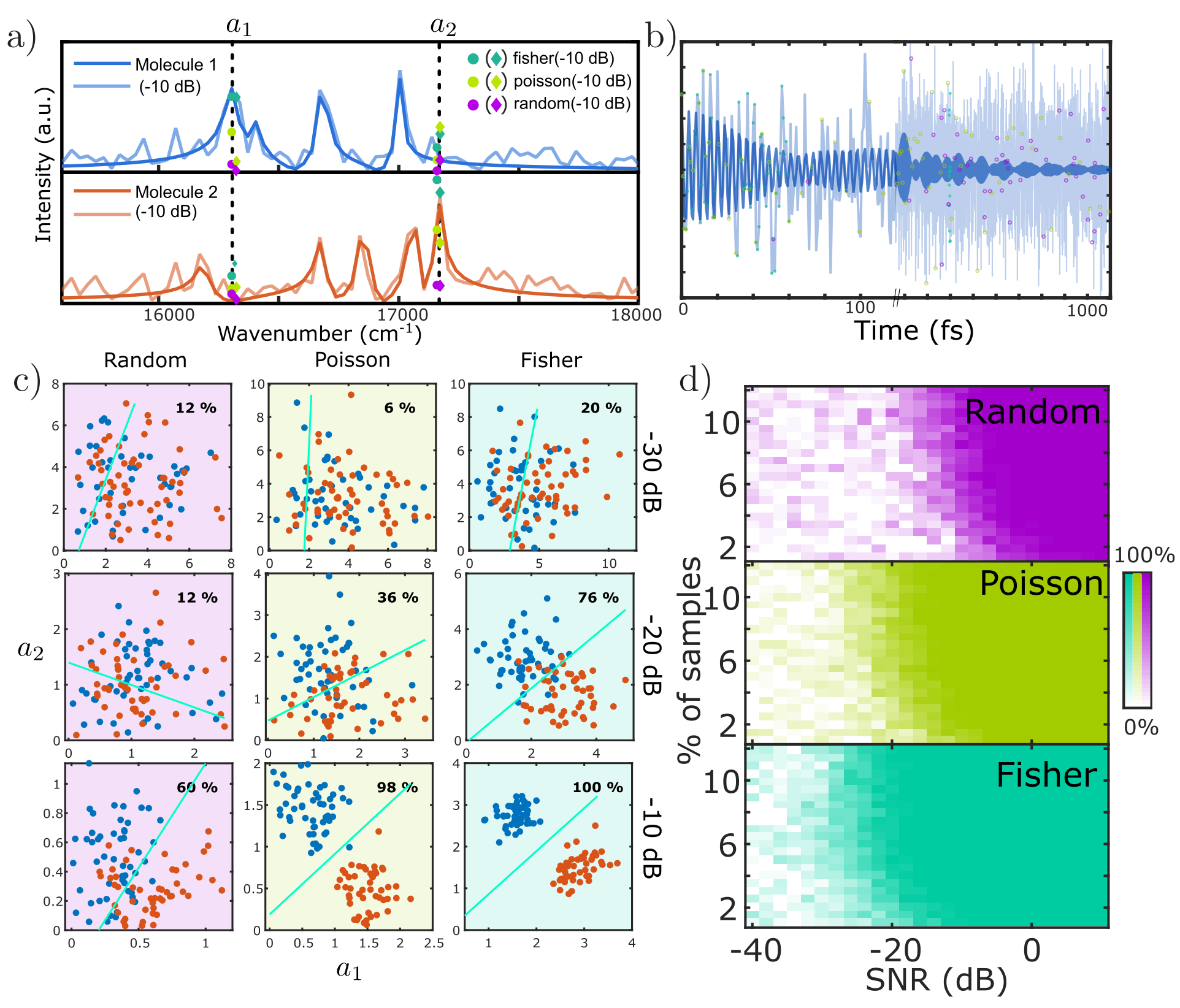}
\caption{a) Simulated frequency domain Raman spectra for two molecules. b) The real part of the time-domain spectrum for molecule 1. The dashed vertical lines indicate the peaks selected for classification, avoiding overlap with the other molecule's peaks. The colored dots in a) correspond to amplitude estimations using LS estimation on the selected data sets shown in b), where the colors represent samples selected via the Fisher, Poisson, and random sampling methods. c) Scatter plots of fitted amplitudes from 100 Monte-Carlo simulations across different SNRs, demonstrating the Fisher method's reduced amplitude variance and improve classification efficiency. d) A detailed comparison of classification accuracy between the Fisher, Poisson, and the random sampling schemes, relative to SNR and data usage percentage.}
\label{fig3}
\end{figure*}

We explore three different sampling methods: the Fisher method, a uniformly distributed random selection, and Poisson gap sampling. The latter, transitioning from denser to sparser sampling, has proven effective in free induction decay data\cite{Kasprzak2021ClusteredSampling}. For practicality and easier visualization, we focus on the amplitudes of just two spectral coordinates for each molecule type, denoted by vertical dashed lines in Figure~\ref{fig3}(a). These coordinates reflects the position of the two peaks that are less overlapping between the two molecules, as this simplifies the distinction between them. 

The Fisher sampling employs a model assuming all ten Lorentzian functions for signal reconstruction, with only the weights of the amplitudes of the two "classification" peaks are left non-zero to focus the minimization on their variances.
A Monte-Carlo simulation is set to choose a molecule, take its signal along the full grid, add normal distributed random noise, picking a number of observation coordinates according to the percentage of data and the method, and estimate the amplitudes of the peaks, $a_1$ and $a_2$, using LS.
The amplitude estimates for each iteration are plotted in scatter plots in Figure~\ref{fig3}c), revealing the variance across different SNRs. Supervised linear discrimination, based on the separation of amplitude sets in the \( [a_1, a_2] \) space, evaluates the classification efficiency. It is noticeable how the fisher method reduces the spread of the amplitudes, that translates into enhanced classification reliability.
Figure~\ref{fig3}d) illustrates the robustness of the Fisher method, which is clearly achieving superior classification efficiency as compared to both the Poisson gap and random sampling methods. As can be seen, the method shows an efficiency reduction at $\approx 5 dB$ lower SNR with respect of the Poisson method and even $\approx 15 dB$ with respect of the random selections, i.e., more than one order of magnitude lower SNR.

\section{Conclusion}

In this work, we present some examples of the application of the Fisher information in spectroscopy. After the definition of the method, we considered a set of case studies, with different functions modeling the expected spectroscopic signal, going from exponential to oscillating samples and from real- to complex-valued acquisitions. We particularly emphasize the differences between this "parameter-centric" method and the standard compressive sampling methods that use a random sampling pattern to reconstruct the signal with high fidelity.
The Fisher method is also shown to be able to visualize where most of the information on the parameters of response functions of 2DES spectra lies, even without precise foreknowledge of the parameters values but just knowing the range or the distributions of the values.
Finally, we demonstrated how this method outperforms other sampling methods in classification of FT-Raman molecular spectra.

\paragraph{Aknowledgment}
The authors would like to acknowledge helpful discussions with Prof.~Tonu Pullerits, Mr.~Edvin Sanden, and Mr.~Pontus Walan.
\\ \\
L.B. received funding from the Clean Planet Program supported by Fundació Joan Ribas Araquistain (FJRA).\\
N.F.v.H. acknowledges the financial support by the European Commission (ERC Advanced Grant 101054846 - FastTrack), QuantERA project no.731473 and 101017733: ExTRaQT), the MCIN/AEI projects PID2021-123814OB-I00, TED2021-129241B-I00, the “Severo Ochoa" program for Centres of Excellence in R\&D CEX2019-000910-S, Fundació Privada Cellex, Fundacio Privada Mir-Puig, and the Generalitat de Catalunya through the CERCA program.\\
A.J. acknowledges the Swedish SRA ESSENCE (grant no. 2020 6:2).
\bibliographystyle{ieeetran}
\bibliography{references}

\end{document}